\documentclass[
    ,final            
  ]
  {aipproc}

\layoutstyle{6x9}

\newcommand{\be}{\begin{equation}}
\newcommand{\ee}{\end{equation}}

\begin{document}

\title{Effect of network topology on the ordering dynamics of voter models}

\author{Claudio Castellano}{
  address={Dipartimento di Fisica, Universit\`a di
Roma ``La Sapienza'' and Center for Statistical Mechanics and Complexity,
INFM Unit\`a Roma 1, P.le A. Moro 2, 00185 Roma, Italy}
}

\begin{abstract}
We introduce and study the reverse voter model, a dynamics for spin
variables similar to the well-known voter dynamics. The difference is
in the way neighbors influence each other:
once a node is selected and one among its neighbors chosen,
the neighbor is made equal to the selected node, while in the usual
voter dynamics the update goes in the opposite direction.
The reverse voter dynamics is studied analytically, showing that
on networks with degree distribution decaying as $k^{-\nu}$,
the time to reach consensus is linear in the system size $N$ for all $\nu>2$.
The consensus time for link-update voter dynamics is computed as well.
We verify the results numerically on a class of uncorrelated scale-free
graphs.

\end{abstract}

\maketitle

\section{Introduction}
Ordering dynamics is a classical topic in statistical physics,
actively investigated for over two decades~\cite{Bray94}.
The best known realization of an ordering process is the dynamics of
a ferromagnet suddenly quenched from a high-temperature state to a
temperature below the critical one: ordered domains are formed and coarsen
in time until the equilibrium state at temperature $T$ is reached.

More recently, ordering processes have started to be investigated
in the completely different context of social sciences: the spreading
of rumors or beliefs, the dynamics of opinions, the diffusion of
cultural traits are all phenomena with the common feature that,
starting from a disordered
initial configuration, the dynamics tends to increase the similarity of
interacting agents, leading toward a more ordered state.
Simple (often oversimplified) models for such phenomena are akin to
models studied in traditional statistical physics, but from the point of view
of social sciences it is more interesting to study their behavior on
networks, rather than on regular lattices.

The voter model is one of the simplest ordering processes~\cite{Liggett85}.
Each individual is fully specified by a single `spin' variable that
can assume only two states, say $-1$ and 1. At each time step an individual
is chosen at random and his variable is set equal to the variable
of a randomly chosen nearest neighbor. This step is then iterated many times.
There are two absorbing configurations for the dynamics:
all variables equal to $-1$ or all equal to $1$. Once these consensus states
are reached no further evolution is possible.
When $N$, the number of agents in the system, is finite, the dynamics
reaches sooner or later one of the two absorbing configurations.
An interesting quantity is then $\tau(N)$, the mean time needed to
arrive at full consensus.
For regular lattices in $d$ dimensions, this time scales as $N^2$ in $d=1$,
$N \ln(N)$ in $d=2$ and as $N$ for $d>2$~\cite{Liggett85, Krapivsky92}.

Some activity has been recently devoted to the study of the
voter model on graphs. In particular, it has been shown that the time
to reach consensus scales as $N$ on the Watts-Strogatz
graph~\cite{Castellano03, Vilone04}.
On the Barabasi-Albert (BA) scale-free network
it is found numerically that $\tau$ grows with $N$ with an exponent
close to $0.88$~\cite{Suchecki04,Castellano05}.
At odds with what occurs on regular lattices, voter dynamics
does not conserve the average magnetization when the number
of nearest neighbors (degree) changes from node to
node~\cite{Suchecki04,Wu04}.
In a very recent contribution~\cite{Sood04}, Sood and Redner have
evaluated analytically the consensus time for a generic uncorrelated
heterogeneous graph with a scale-free degree distribution
$n_k \sim k^{-\nu}$. It turns out that $\tau(N)$ scales as $N$ for $\nu>3$,
it is a constant for $\nu<2$, and it scales as
$N^{(2 \nu-4)/(\nu-1)}$ for $2 < \nu < 3$.
The presence of a logarithmic correction for $\nu=3$, $\tau(N) \sim N/\ln(N)$
accounts well for the observed exponent $0.88$ on the BA network.

In the elementary step of the original voter model, the selected node A
picks at random a nearest neighbor B and becomes equal to it.
When one considers simple modifications of this dynamics there
are two possibilities that immediately come to mind.
The first, similar to the so called Ochrombel simplification
of the Sznajd model~\cite{Ochrombel01},
consists in reversing the direction of the update, so that it is the
neighbor B that becomes equal to the selected node A. We call this
dynamics 'reverse' voter model.
The second is the 'link-update' voter model~\cite{Suchecki04}:
instead of a site, a link is randomly chosen and the update occurs
in random direction along it.
The link-update dynamics conserves the average
magnetization for any degree distribution~\cite{Suchecki04}.
It is clear instead that the 'reverse' voter model violates this conservation.

Whenever the degree distribution is a delta function the three models
perfectly coincide, but this is no longer true on a generic graph.
This is evident for example on a star graph, where all links connect
a hub with the other $N-1$ nodes.
In a single elementary step, the probability that the hub is updated
is $1/N$ for the usual voter model, $1/2$ for the link-update
dynamics and $1-1/N$ for the reverse voter model.

In this paper we investigate for which topology the three types of voter
dynamics give similar
results and when, instead, the ordering processes are qualitatively different.

\section{Analytical Results}

We consider the reverse voter dynamics on a generic heterogeneous
uncorrelated graph with normalized degree distribution $n_k$.
Let us define $\rho_k$, the fraction of up spins on nodes of degree $k$,
and compute the probability $P(k;- \to +)$ that a spin down on a node
of degree $k$ flips in an elementary update.

For this flip to occur it is necessary that the spin on
the selected node (A) is up (and this happens with probability
$\rho = \sum_k n_k \rho_k$) and that the neighbor chosen (B) has degree $k$
and is down.
The degree distribution of the neighbors of a random node in an uncorrelated
network is $k n_k/\mu_1$, where $\mu_i = \sum_k k^i n_k$.
The probability that a node of degree $k$ is down is $1-\rho_k$, hence
\be
P(k;- \to +) = \rho {k n_k \over \mu_1} (1-\rho_k).
\ee
Similarly $P(k;+ \to -) = (1-\rho) {k n_k \over \mu_1} \rho_k$.

Using these expressions one can apply
the analytical treatment of Ref.~\cite{Sood04}
to compute the consensus time $\tau(N)$.
The idea is to write down a recursion formula for $\tau$ as a function of the
$\rho_k$. Expanding to second order one obtains
\be
\sum_k {k \over \mu_1} (\rho_k-\rho) \partial_k \tau -
{1 \over 2N} \sum_k {k \over \mu_1 n_k} (\rho + \rho_k -2 \rho \rho_k)
\partial^2_k \tau = 1,
\label{eqrec}
\ee
where $\partial_k = \partial/\partial \rho_k$.

To analyze the role of the different terms in Eq.~(\ref{eqrec}) we have
to consider the dynamics of the densities $\rho_k$. In an elementary
update, the number $N_k^{up}$ of up spins of degree $k$, changes by
an amount $d N_k^{up} = P(k;- \to +) - P(k;+ \to -)$,
so that 
\be
\dot{\rho_k} = {d N_k^{up} \over N_k dt} =
{1 \over n_k N dt} [P(k;- \to +)-P(k;+ \to -)]  = {k \over \mu_1}(\rho-\rho_k).
\ee

The equation of motion for the magnetization $\rho$ is
\be
\dot{\rho} = \rho- {1 \over \mu_1} \sum_k n_k k \rho_k.
\ee

Starting from their initial values, the densities $\rho_k$ rapidly converge
to a stationary value
\be
\rho_k = \rho_s~~~~~~~~~~~~\forall k.
\label{rho_s}
\ee

This stationary value can be related to the quantity
$R=\mu_1 \sum_k {n_k \over k} \rho_k$ that is conserved by the dynamics
\be
R = \mu_1 \rho_s \sum_k {n_k \over k} \rho_k = {\rho_s \over C},
\label{R}
\ee
where $C=1/(\mu_1 \mu_{-1})$.

It is now possible to rewrite Eq.~(\ref{eqrec}) in the stationary state.
Using Eqs.~(\ref{rho_s}) and~(\ref{R}) the first term is seen to vanish;
one is left with
\be
R (1-C R) \partial^2_R \tau = -N.
\label{eqR}
\ee
The integration of Eq.~(\ref{eqR}), with boundary conditions
$\tau(R=0)=0$ and $\tau(R=1/C)=0$, yields
\be
\tau(N) = - N \left[R \ln(CR) + \left({1 \over C}-R \right) \ln(1-C R)\right].
\ee

For an uncorrelated initial configuration, i.e. $\rho_k(0)=\rho(0)~~\forall k$,
\be
\tau(N) = - N \mu_1 \mu_{-1} \left[\rho(0) \ln \rho(0) + (1 - \rho(0))
\ln(1-\rho(0)) \right].
\label{tau_o}
\ee

Since the prefactor is finite the time to consensus
diverges linearly with $N \to \infty$  for all $\nu>2$.
For $\nu=2$ a logarithmic correction is present: $\tau(N) \propto N \ln(N)$,
while $\tau(N) \sim N^{1/(\nu-1)}$ for $1 <\nu <2$.

Applying the same analytical approach, the consensus time for link-update
dynamics~\cite{Suchecki04} can be computed as well.
The probability of flipping a down spin on a node of degree $k$ has the form
$P(k;- \to +) = \omega {k n_k \over \mu_1} (1-\rho_k)$, where
$\omega= \sum_k n_k k \rho_k$. The consensus time turns out to be
\be
\tau(N) = -N [(1-\rho) \ln (1-\rho) + \rho \ln \rho].
\label{tau_lu}
\ee
Also in this case the time to consensus depends linearly on the
system size $N$. Remarkably, there is no dependence at all on the
network parameters.

\section{Numerical Results}

In order to validate the anaytical results we have
performed numerical simulations of the reverse voter dynamics on two
classes of scale-free graphs. 

\begin{figure}
  \includegraphics[height=.3\textheight]{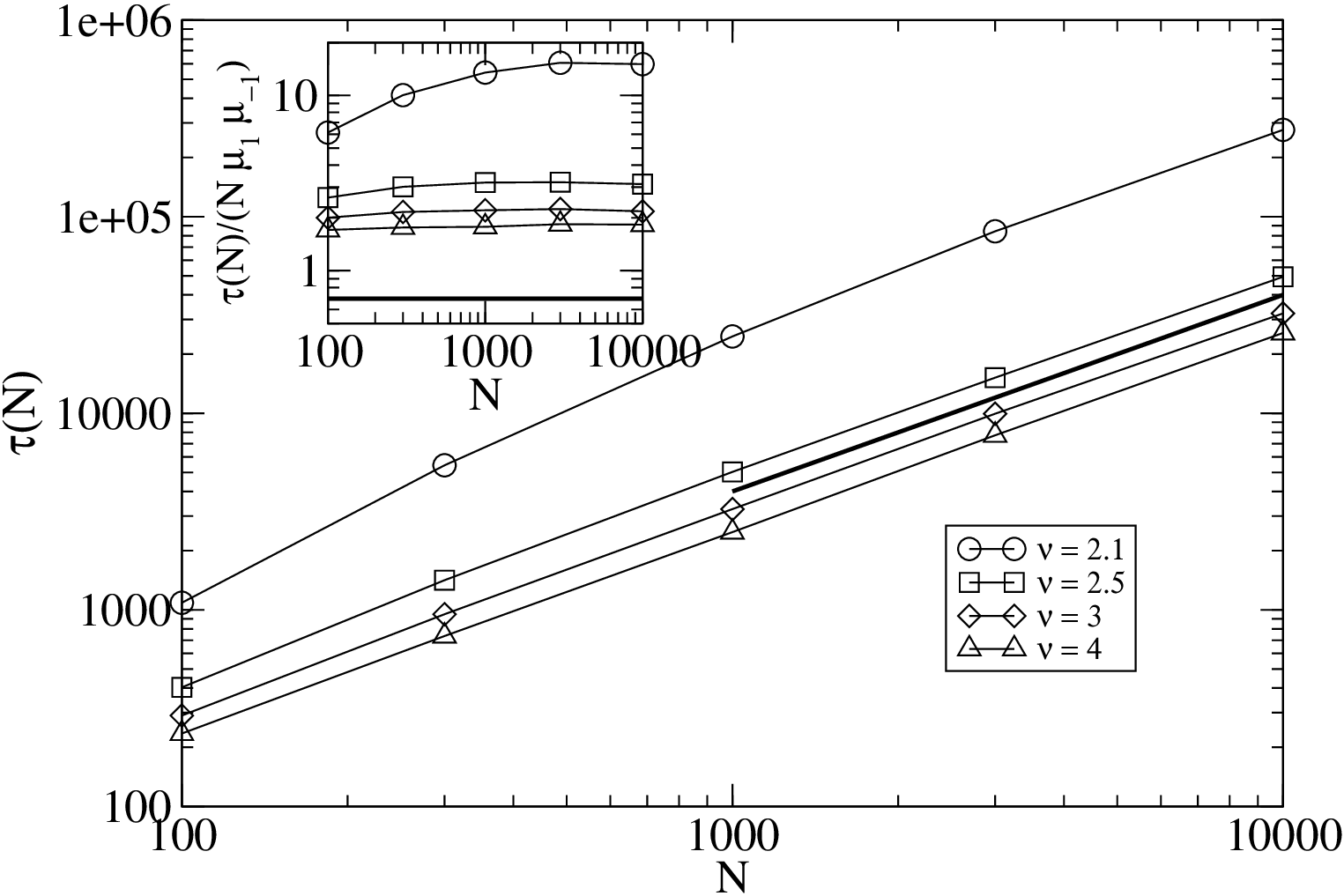}
  \caption{Main: Double logarithmic plot of the consensus time $\tau(N)$
as a function of $N$ for the reverse voter dynamics on a growing network
with redirection. The thick line has slope 1. Inset: The same data divided
by $N \mu_1 \mu_{-1}$. The thick line is here the result predicted by 
Eq.~(\ref{tau_o}).}
\label{Fig1}
\end{figure}

In Fig.~\ref{Fig1} we present the consensus time for reverse voter dynamics
on a growing network with redirection~\cite{Krapivsky01} for several values
of the exponent $\nu$ of the scale-free degree distribution.

From the main part of the figure it is clear that, while the asymptotic
behavior is, as expected, proportional to the system size, sizeable
corrections occur for smaller values of $N$.
This might be caused by the residual preasymptotic dependence
of $\mu_1$ and $\mu_{-1}$ on $N$, for the relatively small values of
$N$ considered.
In the inset we have therefore plotted $\tau(N)/(N \mu_1 \mu_{-1})$
and compared to the value $\ln(2)=0.6931...$, predicted by Eq.~(\ref{tau_o}).
The agreement is not satisfactory. This suggests that correlations
present in the network have an important influence on the ordering
process; they do not alter the asymptotic linear dependence on $N$
but affect rather strongly the prefactor.

We have then considered the evolution of the reverse voter model on
another class of graphs, recently introduced by Catanzaro et
al.~\cite{Catanzaro05}.
Such graphs are built according to the usual algorithm for the
configuration model, with the additional prescription that no node can
have a degree larger than $N^{1/2}$.
This modification has the goal of avoiding the correlations
of the node connectivities present in the usual configuration model.

\begin{figure}
  \includegraphics[height=.3\textheight]{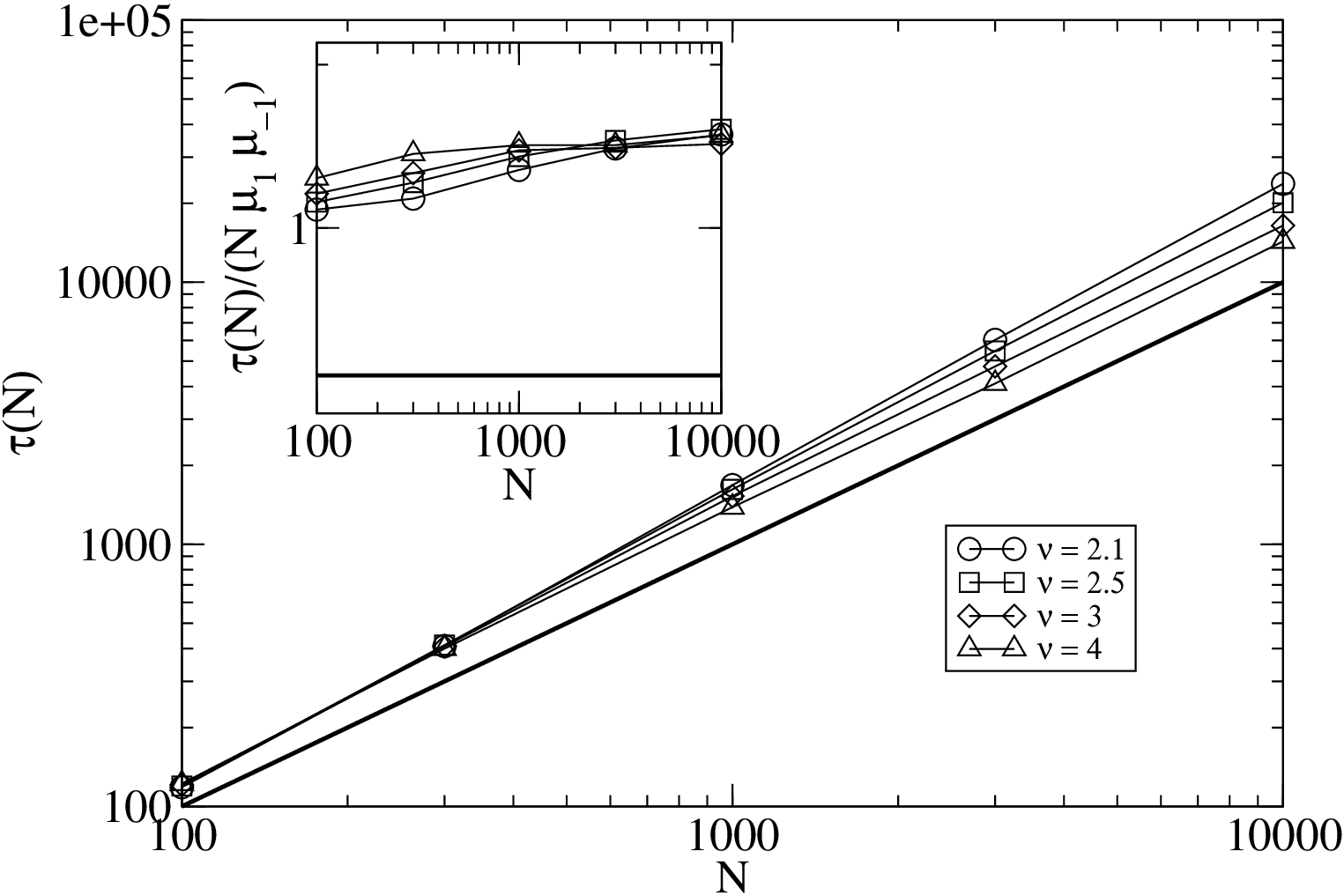}
  \caption{Main: Double logarithmic plot of the consensus time $\tau(N)$
as a function of $N$ for the reverse voter dynamics on the
uncorrelated scale-free networks of Ref.~\cite{Catanzaro05},with minimum
degree equal to 3, for $\rho_k(0)=\rho(0)=1/2$.
The thick line has slope 1. Inset: The same data divided
by $\mu_1 \mu_{-1}$. Here the thick line is the result predicted by 
Eq.~(\ref{tau_o}), $\ln(2)$.}
\label{Fig2}
\end{figure}

It is evident from the results, displayed in Fig.~\ref{Fig2}, that
even on this class of networks one does not see a clean linear
dependence on $N$ for the system sizes considered here.
The effective exponent for small $N$ is larger than one; a crossover
to the expected asymptotic behavior is seen for larger values of $N$.
The plot of $\tau(N)/(N \mu_1 \mu_{-1})$, presented in the inset,
indicates that the bending in the main plot actually comes from the
residual $N$-dependence of $\mu_1$ and $\mu_{-1}$.
The quantity $\tau(N)/(N \mu_1 \mu_{-1})$ goes asimptotically to
a constant value that, as predicted from Eq.~(\ref{tau_o}) does not
depend on the exponent $\nu$.
The only feature of Fig.~\ref{Fig2} not in agreement with the theory
is the value of the constant, that is actually of the order of two times
the expected value, $\ln(2)$.

\begin{figure}
  \includegraphics[height=.3\textheight]{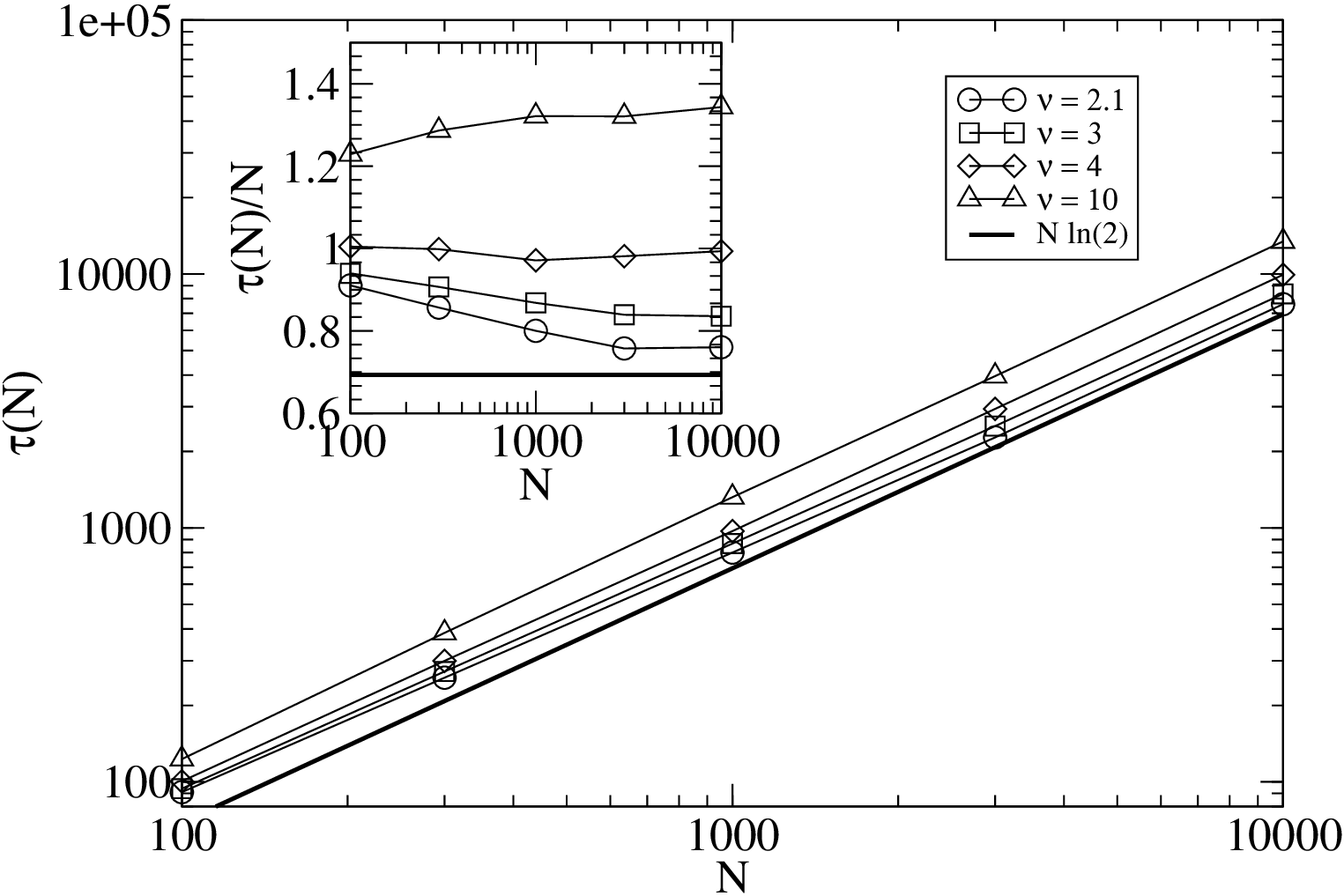}
  \caption{Main: Double logarithmic plot of the consensus time $\tau(N)$
as a function of $N$ for the link-update voter dynamics on the
uncorrelated scale-free networks of Ref.~\cite{Catanzaro05} with minimum
degree 3, for $\rho_k(0)=\rho(0)=1/2$.
Inset: The same data divided  by $N$. The thick line is the result predicted by
Eq.~(\ref{tau_lu}), $\ln(2)$.}
  \label{Fig3}
\end{figure}

On the uncorrelated scale-free graphs we have also studied numerically
the ordering of the link-update voter model.
Although Eq.~(\ref{tau_lu}) predicts the consensus time to be completely
independent from $\nu$ we find that the curves do not superimpose
(Fig.~\ref{Fig3}).
A simple linear fit yields exponents ranging from 0.96 to 1.02, but
a more careful inspection indicates that the curves are slightly bent.

Further insight is provided by the inset, where the consensus time
divided by the system size $N$ is plotted as a function of $N$.
A transient regime, for small $N$, with non constant $\tau(N)/N$
is followed by plateau, indicating that the asymptotic linear regime
is reached.
At odds with Eq.~(\ref{tau_lu}),
the height of the plateau depends on the value of $\nu$, ranging from
approximately one to two times the expected value $\ln(2)$.

Hence we find that also in this case the analytical agreement
correctly captures the linear behavior independent from
$\nu$, while it fails for what concerns the prefactor.
We do not have a clear understanding of the origin of this disagreement.
One possible reason could be the presence of residual correlations in
the modified configuration model that we have used.
However, the mismatch occurs also for values of $\nu$ so large that
correlations are surely negligible.
It is more likely that the different prefactors are an effect of the
term containing the first derivative in Eq.~(\ref{eqrec}), that we
have neglected in the analytical treatment, but that may play
a role in the initial stage of the ordering process.

\section{Conclusions}

We have studied analytically and numerically the ordering behavior of the
reverse and the link-update voter models, finding that the time to
reach consensus grows linearly with the system size for scale-free networks
with $\nu>2$.

It is interesting to notice that, for the reverse voter model, the time
to reach consensus gets larger as $\nu$ is decreased, since $\mu_1$ grows.
This can be qualitatively understood as follows: in the reverse voter dynamics
it is very easy that a hub changes its state, being chosen as the neighbor
of the selected node.
In the usual voter dynamics, instead, the presence of a hub 
favors order, since its many neighbors tend to become equal to it.
Correspondingly, the time to consensus in the normal voter dynamics
is reduced as $\nu$ is decreased~\cite{Sood04}.

\begin{figure}
  \includegraphics[height=.3\textheight]{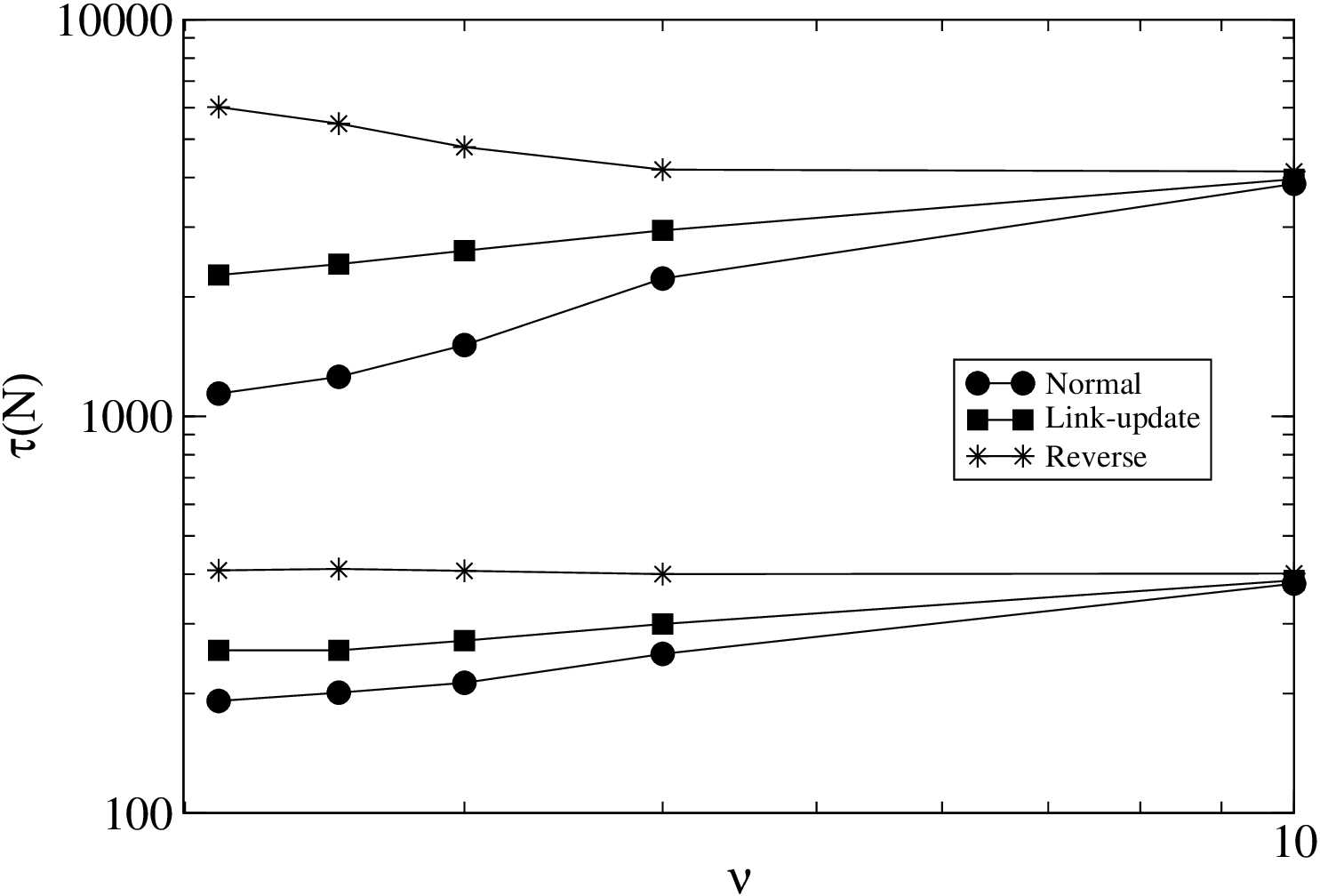}
  \caption{Main: Plot of the consensus time $\tau$
as a function of $\nu$ for the three types of voter dynamics on the
uncorrelated scale-free networks of Ref.~\cite{Catanzaro05} with minimum
degree 3, for $\rho_k(0)=\rho(0)=1/2$. The upper curves are for $N=3000$,
the lower curves are for $N=300$.
}
  \label{Fig4}
\end{figure}

As mentioned above, the three types of voter model are perfectly equal
when the degree is the same for all nodes.
This can be seen also from Eqs.~(\ref{tau_o}) and~(\ref{tau_lu}).
For $\nu \to \infty$, $\mu_i = (\mu_1)^i$ and in all cases (including
the usual voter dynamics), the consensus time is given by Eq.~(\ref{tau_lu}).
For finite values of $\nu$, instead, the usual voter dynamics leads
to consensus most quickly, and the reverse dynamics is the
slowest~(Fig.~\ref{Fig4}).

All the results presented here were obtained under the assumption
that the dynamics takes place on an uncorrelated network.
An interesting issue for future work is the investigation of voter
dynamics on correlated graphs.



\bibliographystyle{aipproc}   

\bibliography{sample}

\IfFileExists{\jobname.bbl}{}
 {\typeout{}
  \typeout{******************************************}
  \typeout{** Please run "bibtex \jobname" to optain}
  \typeout{** the bibliography and then re-run LaTeX}
  \typeout{** twice to fix the references!}
  \typeout{******************************************}
  \typeout{}
 }

\end{document}